\documentclass{jfm}
\usepackage{graphicx}
\usepackage{natbib}

\newcommand{\del}{\mbox{\boldmath{$\nabla$}}}

\newcommand{\nhat}{\mbox{\boldmath{$\hat{n}$}}}

\newcommand{\xhat}{\mbox{\boldmath{$\hat{x}$}}}
\newcommand{\yhat}{\mbox{\boldmath{$\hat{y}$}}}
\newcommand{\zhat}{\mbox{\boldmath{$\hat{z}$}}}
\newcommand{\ehat}{\mbox{\boldmath{$\hat{e}$}}}
\newcommand{\ub}{\mbox{{\bf u}}}
\newcommand{\xb}{\mbox{{\bf x}}}
\newcommand{\Ub}{\mbox{{\bf U}}}
\newcommand{\vb}{\mbox{{\bf v}}}

\newcommand{\gb}{\mbox{{\bf g}}}
\newcommand{\bb}{\mbox{{\bf b}}}
\newcommand{\Ab}{\mbox{{\bf A}}}
\newcommand{\Bb}{\mbox{{\bf B}}}
\newcommand{\Cb}{\mbox{{\bf C}}}
\newcommand{\Db}{\mbox{{\bf D}}}
\newcommand{\Mb}{\mbox{{\bf M}}}
\newcommand{\Fb}{\mbox{{\bf F}}}
\newcommand{\Ib}{\mbox{{\bf I}}}
\newcommand{\Ob}{\mbox{{\bf 0}}}
\newcommand{\Sb}{\mbox{{\bf S}}}
\newcommand{\fb}{\mbox{{\bf f}}}
\newcommand{\Kb}{\mbox{{\boldmath{$K$}}}}
\newcommand{\sigmab}{\mbox{{\boldmath{$\sigma$}}}}
\newcommand{\kappab}{\mbox{{\boldmath{$\kappa$}}}}

\begin{document}

\title[Tensorial hydrodynamic slip]{Tensorial hydrodynamic slip} 

\author[M. Z. Bazant and
O. I. Vinogradova]{M\ls A\ls R\ls T\ls I\ls N\ns Z.\ns B\ls A\ls Z\ls
  A\ls N\ls T$^{1,3}$ and O\ls L\ls G\ls A\ns I.\ns V\ls I\ls N\ls
  O\ls G\ls R\ls A\ls D\ls O\ls V\ls A$^{2,3}$\ \\} 

\affiliation{$^1$  Department of Mathematics,
  Massachusetts Institute of Technology, Cambridge, MA 02139\\
  $^2$ A.N.~Frumkin Institute of Physical Chemistry and
  Electrochemistry, Russian Academy of Sciences, 31 Leninsky
  Prospect, 119991 Moscow, Russia\\
  $^3$ CNRS UMR Gulliver 7083 and 7636, \'Ecole Sup\'erieure de Physique et de
  Chimie Industrielles, 10 rue Vauquelin, F-75005 Paris, France}
\setcounter{page}{1} \date{\today}
\label{firstpage}
\maketitle

\begin{abstract}
  We describe a tensorial generalization of the Navier slip boundary
  condition and illustrate its use in solving for flows around
  anisotropic textured surfaces. Tensorial slip can be derived from
  molecular or microstructural theories or simply postulated as an
  constitutive relation, subject to certain general constraints on the
  interfacial mobility. The power of the tensor formalism is to
  capture complicated effects of surface anisotropy, while preserving
  a simple fluid domain. This is demonstrated by exact solutions for
  laminar shear flow and pressure-driven flow between parallel plates
  of arbitrary and different textures. From such solutions, the
  effects of rotating a texture follow from simple matrix algebra. Our
  results may be useful to extracting local slip tensors from global
  measurements, such as the permeability of a textured channel or the
  force required to move a patterned surface, in experiments or
  simulations.
\end{abstract}

\section{ Introduction}

The emergence of microfluidics has focused renewed attention on
hydrodynamic boundary conditions~\citep{stone2004}. Reducing fluid
volumes enhances the impact of surface phenomena~\citep{squires2005},
so the use of appropriate boundary conditions is crucial to the design
and optimization of lab-on-a-chip devices. It is now widely recognized
that the classical no-slip hypothesis supported by macroscopic
experiments does not always apply at the micro- and, especially,
the nano-scale.

In this context, the phenomenon of liquid slip at solid surfaces has
been studied extensively in experiments, theoretical calculations, and
simulations~\citep{vinogradova1999,lauga2005,bocquet2007}.  The
results are usually interpreted in terms of the Navier boundary
condition,
\begin{equation}
\Delta u = u - U = b \, \frac{\partial u}{\partial n} \label{eq:navier}
\end{equation}
where the fluid velocity $u$ minus the surface velocity $U$ is
proportional to the shear strain rate $\partial u/\partial n$ via the
slip length $b$. Flow past smooth hydrophilic surfaces has been shown
to be consistent with the no-slip hypothesis, but $b$ can reach tens
of nanometres for hydrophobic
surfaces~\citep{vinogradova2003,charlaix.e:2005,joly.l:2006}.
Hydrophobicity can be significantly amplified by roughness and can
reduce friction due to trapped
nanobubbles~\citep{vinogradova.oi:1995b,cottin_bizonne.c:2003.a}. Extreme
hydrophobicity can be generated with well-controlled
textures~\citep{quere2005}, leading to a many-micron slip
lengths~\citep{ou2005,joseph.p:2006,choi.ch:2006} and very fast
transport of water through microchannels. The strong anisotropy of
such surfaces, however, can limit the validity of
Eq.~(\ref{eq:navier}).

The possibility of transverse flow over a grooved no-slip surface,
perpendicular to an applied shear stress, has been analyzed by
\cite{stroock2002b}, ~\cite{ajdari2002}, and \cite{wang2003} and
exploited for chaotic mixing in microfluidic devices by
~\cite{stroock2002a}. In this context, \cite{stroock2002b} expressed
the permeability $\kappab$ of a thin (parallel-plate) microchannel
with one grooved and one flat surface in terms of an effective {\it
  slip-length tensor}, $\bb = \{ b_{ij} \}$, defined by a generalized
Navier boundary condition
\begin{equation}
\Delta \ub = \ub - \Ub = \bb\, (\nhat \cdot \del \ub)  \label{eq:tensorbc}
\end{equation}
and \cite{stone2004} expressed the velocity profile in terms of
$\bb$. This elegant construction relating permeability to slip,
however, assumes that the {\it global} flow has the same anisotropy as
the grooved surface (i.e.  $\kappab$ and $\bb$ are coaxial). This is
generally not the case with multiple textured
surfaces~\citep{wang2003}, curved walls~\citep{einzel.d:1990},
obstacles in the flow, etc., and we are not aware of any other use of the
tensorial relation (\ref{eq:tensorbc}). Notably, \cite{wang2003}
considered flow between misaligned, grooved plates using
(\ref{eq:tensorbc}) in component form but deemed the solution `too
tedious to reproduce'. We shall see that this problem and others have
very simple solutions in tensorial form.

In this article, we propose the use of ~(\ref{eq:tensorbc}) as a {\it
  local} boundary condition for any surface whose texture perturbs
fluid flow on length scales much smaller than the geometry. We begin
in section ~\ref{sec:bc} by discussing a general boundary condition
relating slip velocity to normal traction via an interfacial mobility
tensor. To illustrate its use, we derive exact solutions for two types
of laminar flow between textured parallel plates (which can also be
superimposed): (i) shear flow due to moving plates in section
~\ref{sec:shear}, and (ii) pressure-driven flow in section
~\ref{sec:pois}.  We close in section~\ref{sec:disc} by suggesting
further applications.

\section{ Theory }
\label{sec:bc}

\subsection{ The interfacial mobility tensor }

Although Equation (\ref{eq:navier}) is the most commonly used boundary
condition for hydrodynamic slip, it is not widely appreciated that
\cite{navier1823} also postulated the more general relation,
\begin{equation}
\Delta u = M \tau \label{eq:navier2}
\end{equation}
where $\tau$ is the local shear stress (normal traction) and $M$ is a
constant interfacial mobility (velocity per surface stress). For a
Newtonian fluid, $\tau = \eta \partial u /\partial n$, this reduces to
(\ref{eq:navier}) with $b = M\eta$, where $\eta$ is the
viscosity. Molecular dynamics simulations have shown that
(\ref{eq:navier2}) with constant $M$ is more robust than
(\ref{eq:navier}) with constant $b$, since the fluctuating slip
velocity correlates better with the shear stress (normal forces) than
with velocity gradients very close to the surface
\citep{hess1989,bocquet2007}.

\begin{figure}
\begin{center}
\vspace{0.1in}
\includegraphics[width=3.3in]{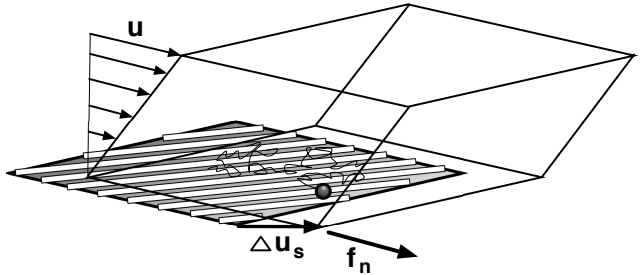}
\caption{ \label{fig:slab} Physical picture of tensorial slip. The
  normal traction $\fb_n$ exerted by the fluid on an anisotropic
  surface produces an effective slip velocity $\Delta\ub=\Mb \, \fb_n$
  in a different direction. At the molecular level, the interfacial
  mobility tensor $\Mb$ can be related to trajectories of diffusing
  interfacial particles, such as the one shown.  }
\end{center}
\end{figure}

A natural generalization of the slip condition (\ref{eq:navier2}) is
\begin{equation}
\Delta \ub  = \Mb \, (\nhat \cdot \sigmab) \label{eq:M}
\end{equation}
where $\nhat \cdot \sigmab = \fb_n$ is the fluid force (normal
traction) on the interface, $\sigmab$ is the local stress tensor, and
$\Mb$ is an interfacial mobility tensor. As shown in
Figure~\ref{fig:slab}, the effective slip vector is generally
misaligned with the force vector for an anisotropic surface.  Equation
(\ref{eq:tensorbc}) is recovered with $\bb = \Mb \eta$ in the case of
a Newtonian fluid of viscosity $\eta$.  For anisotropic surfaces, the
mobility is a second-rank tensor $\Mb = \{M_{ij}\}$, whether the
averaging of surface forces occurs over microstructural or
molecular heterogeneity.

As with scalar slip~\citep{bocquet2007}, the tensorial slip boundary
condition (\ref{eq:navier2}) can be justified in various ways.  At the
microstructural level, grooved surfaces (with or without scalar slip)
have effective tensorial slip coefficients, which can be explicitly
calculated for simple geometries, if the grooves vary on much smaller
scales than the fluid domain~\citep{stroock2002b,wang2003}. At the
molecular level, nanoscale surface anisotropy has a similar effect,
but due to statistical interactions.

A possible starting point for molecular modeling is a tensorial
Einstein relation, $\Db = \Mb kT/S$, relating the $\Mb$ to the
`interfacial diffusivity' per unit area $S$, by analogy with the
theory of Brownian motion. This yields the statistical formula
\begin{equation}
  M_{ij} = \frac{S}{2kT} \lim_{t\to\infty} \frac{d}{dt} \, 
  \mbox{Cov}(\Delta x_i(t), \Delta x_j(t))  \label{eq:dif}
\end{equation}
where $\Delta \xb(t)=\xb(t) - \Ub t$ is the fluctuating position of an
interfacial fluid molecule, in a frame moving with the mean surface
velocity (Fig.~\ref{fig:slab}), where the `interface' may include
molecules distinct from the bulk fluid, e.g. in a vapor phase.  The
idea of observing thermal diffusion near a surface to infer its slip
length has also been exploited in experiments by \cite{joly.l:2006}.
The mobility formula (\ref{eq:dif}) can also be recast in a tensorial
Green-Kubo form,
\begin{equation}
  M_{ij} = \frac{S}{kT} \int_0^\infty dt \, \mbox{Cov}(v_i(0),v_j(t))
\end{equation}
where $\vb(t) = d\Delta\xb/dt$. These formal expressions assume
convergence in the thermodynamic limit (taken before $t \to \infty$)
for molecular trajectories exploring the interfacial region on scales
much larger than the surface heterogeneity. In that case, via the
covariance matrix, $\Mb$ is symmetric, positive definite and thus
invertible. As noted by ~\cite{bocquet2007}, the inverse mobility, or
friction tensor $\Fb=\Mb^{-1}$, also has a tensorial Green-Kubo
representation, as the integral of the auto-correlation function for
forces exerted by the fluid on the surface~\citep{bocquet1994}. 

\subsection{ General properties of $\Mb$ }

Regardless of its microscopic justification, we suggest adopting
(\ref{eq:M}) as a general, interfacial constitutive relation for
continuum mechanics. As with its bulk counterpart relating the stress
and deformation rate, its form can be either derived from microscopic
models or simply postulated and fit to experimental data, subject to
certain constraints discussed below.  For a general `nonlinear
interface', the mobility tensor $\Mb$ could depend on the surface
forces, as well as internal degrees of freedom, such as the local
orientation of surface molecules or deformable microstructures; for
example, hinge-like structures could lead to different slip in
opposite directions. For permeable surfaces with $\nhat \cdot \Delta
\ub \neq 0$, the mobility tensor may be represented by a $3 \times 3$
matrix with tangential-normal couplings, a possibility which has not
been considered before to our knowledge.

Here, we will focus on the simplest case of impermeable,
macroscopically homogeneous, linear interfaces, where $\Mb$ is a
constant $2\times 2$ matrix in a local suitable coordinate system of
the tangent plane.  Below we will refer to the mobility tensor as
defining the `texture' of a surface up to a rotation, which sets the
`orientation'. The tensor formalism allows us to easily change the
orientation of a texture, once a problem has been solved in terms of
mobility tensors for a given geometry. The mobility simply transforms
as
\begin{equation}
\Mb \mapsto \Sb_\theta\, \Mb \, \Sb_{-\theta} \ \ \mbox{ where } \ \ 
S_\theta = \left( \begin{array}{cc}
\cos\theta & \sin\theta \\
-\sin\theta & \cos\theta \end{array} \right)
\end{equation}
is a matrix rotating the tangent plane by an angle $\theta$.

We also consider `passive' surfaces, which do not transfer energy to
the fluid. In that case, enforcing a positive rate of work $w_I$ on
the slipping interface ($\Mb \neq \Ob$),
\begin{equation}
w_I = \fb_n \cdot \Delta \ub = \fb_n \cdot \Mb \, \fb_n > 0
\end{equation}
for any loading $\fb_n=\nhat\cdot\sigmab$ implies that $\Mb$ must be
positive definite. This argument is similar to the constraint of
positive entropy production at a slipping boundary in irreversible
thermodynamics~\citep{heidenreich2007}. The statistical arguments
above lead to the same conclusion, e.g. since the diffusivity $\Db$ is
positive definite for a passive surface.  The eigenvectors of $\Mb$
correspond to special directions along which fluid forces do not
produce transverse slip, and the (positive) eigenvalues are the
corresponding directional mobilities.  Since positive definite
matrices are invertible, the boundary condition can also be expressed
as $\fb_n = \Fb \, \Delta \ub$ in terms of the (coaxial) friction
tensor, $\Fb=\Mb^{-1}$.

Diagonalization allows us to relate $\Mb$ to the position of the slip
plane in (\ref{eq:M}), which is independent of the force 
$\fb_n$. In each eigendirection $\ehat_i$, the tensorial boundary
condition (\ref{eq:M}) reduces to the scalar case (\ref{eq:navier2}),
and the eigenvalue $M_i$ depends on the (arbitrary) choice of slip
plane in the usual way~\citep{bocquet2007}; for a Newtonian fluid
(\ref{eq:navier}), the slip-length $b_i=M_i\eta$ is the position of
the slip plane, relative to the (unique) depth of no slip extrapolated
from a homogeneous bulk shear flow. By appropriately shifting the
eigenvalues $\{M_i\}$, the same slip plane can be chosen for all
directions. The mobility tensor is then constructed from the spectral
decomposition, $\Mb = \Sb \hat{\Mb} \Sb^{-1}$, where $\hat{\Mb}$ is
the diagonal matrix of eigenvalues and $\Sb$ the matrix
of column eigenvectors.

\subsection{ Symmetric mobility tensors }

Although we will make no further assumptions in our analysis below, a
constant mobility tensor is usually symmetric, $M_{ij} = M_{ji}$, as
in the statistical formulae above. This is also the case for the
effective slip tensor derived by averaging linear Stokes flows over
grooved no-slip surfaces~\citep{stroock2002a,wang2003}. More
generally, symmetry of $\Mb$ exemplifies the widely used
Onsager-Casimir relations of linear response near thermal
equilibrium~\citep{bocquet1994,ajdari2002,heidenreich2007}.

A $2\times 2$ interfacial mobility matrix, which is symmetric and
positive definite, has some useful mathematical properties. There
always exists a rotation of the orthogonal $(x,y)$ coordinate system
of the tangent plane $\Sb_\theta$, which diagonalizes the mobility,
\begin{equation}
  \Mb = \Sb_\theta \left( \begin{array}{cc}
      M_\| & 0 \\
      0 & M_\perp \end{array} \right) \Sb_{-\theta} =
\left( \begin{array}{cc}
M_\| \cos^2 \theta + M_\perp\sin^2\theta & (M_\| - M_\perp)\sin\theta
\cos\theta \\
 (M_\| - M_\perp)\sin\theta
\cos\theta & M_\| \sin^2\theta + M_\perp \cos^2 \theta
\end{array} \right)    \label{eq:Msym}
\end{equation}
where $M_\| \geq M_\perp > 0$ are the eigenvalues.  The decomposition
$\Mb = M_\| \ehat_\| \ehat_\|^T + M_\perp \ehat_\perp \ehat_\perp^T$
shows that $\Delta \ub$ is a linear superposition of scalar slip in
the eigendirections.

Regardless of the complexity of the texture, as long as Equation
(\ref{eq:M}) holds at the geometrical scale with a symmetric, positive
definite $\Mb$, there exist orthogonal directions on the surface,
$\ehat_\| = \Sb_\theta \xhat$ and $\ehat_\perp = \Sb_\theta \yhat$,
along which there are no transverse hydrodynamic couplings,
$\ehat_\perp \cdot \Mb \ehat_\| = 0$. The mobility for `forward' slip
aligned with forcing in a particular direction $\ehat = \Sb_\phi
\xhat$ is given by $ \ehat \cdot \Mb \ehat = M_\| \cos^2(\theta-\phi)
+ M_\perp \sin^2(\theta-\phi)$ and is bounded by the eigenvalues,
$M_\perp \leq \ehat \cdot \Mb \ehat \leq M_\|$. The `fast' axis of
greatest forward slip ($\theta=0$) is always perpendicular to the
`slow' axis of least forward slip ($\theta=\pi/2$).

\subsection{ Simple examples }

In the following sections, we focus on passive linear interfaces and
Newtonian fluids, described by the Navier-Stokes equations
\begin{equation}
\rho \left( \frac{\partial \ub}{\partial t} + \ub \cdot \del \ub
\right) = - \del p + \eta \nabla^2 \ub \ \ \mbox{ and  } \ \ 
\del\cdot\ub = 0   \label{eq:NS}
\end{equation}
In that case, all the properties of $\Mb$ above are inherited by the
slip-length tensor, $\bb = \Mb \eta$, with eigenvalues, $b_\| = M_\|
\eta$ and $b_\perp = M_\perp \eta$. We also assume impermeable,
macroscopically homogeneous surfaces, for which $\bb$ is a constant
$2\times 2$ matrix.

\begin{figure}
\begin{center}
\vspace{0.1in}
\includegraphics[width=4in]{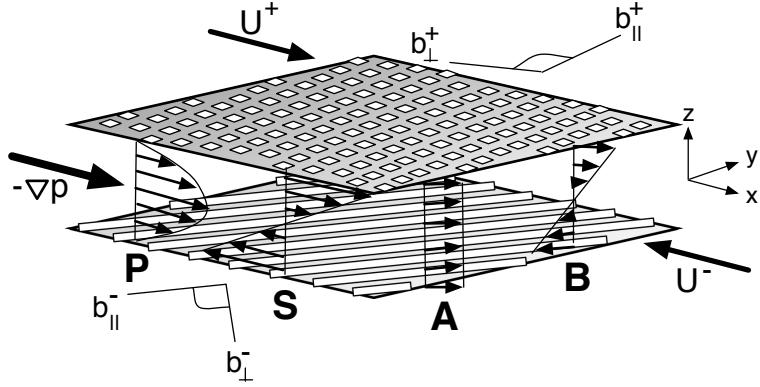}
\caption{ \label{fig:plates} Sketch of a fluid region $|z|< h/2$
  between upper ($+$) and lower ($-$) parallel plates with arbitrary
  textures (grooves, surface coatings, bubbles, etc.). The texture
  length scales are much less than the gap $h$, so each surface has a
  well defined slip-length tensor, $\bb^+$ and $\bb^-$, with
  eigenvalues $b^\pm_\|$ (and $b^\pm_\perp$) in the fastest (and
  slowest) slipping directions indicated. The plates move at relative
  velocity $\Ub$ and/or a uniform pressure gradient $-\del p $ is
  applied. In addition to the usual no-slip parabolic Poiseuille flow
  {\bf P} and linear shear flow {\bf S}, there are superimposed
  slip-driven plug-flow {\bf A} and shear-flow {\bf B}  in
  different directions.  }
\end{center}
\end{figure}

To illustrate the use of slip tensors, we consider the geometry in
Fig.~\ref{fig:plates} where the fluid is confined between flat plates
at $z=\pm h/2$ moving at velocities $\Ub^\pm$ (this section) or forced
by a pressure gradient (next section). Each plate has a fine texture
(varying on scales $\ll h$) and exhibits {\it uniform} tensorial slip,
\begin{equation}
  \ub = \Ub^\pm  \mp \bb^\pm \frac{\partial \ub}{\partial z} \ \ \mbox{ for } \ \
  z = \pm \frac{h}{2}   \label{eq:BC}
\end{equation}
where the slip-length tensors, $\bb^+$ and $\bb^-$, are represented by
constant, positive definite (but not necessarily symmetric) $2\times
2$ matrices in the $(x,y)$ coordinate system.

\section{ Example: Shear flow }
\label{sec:shear}

\subsection{ General solution }

The simplest solution of (\ref{eq:NS})-(\ref{eq:BC}) corresponds to
laminar shear flow between two moving textured plates, shown in
Fig.~\ref{fig:plates}.
In terms of the depth-averaged velocity $\overline{\Ub} = (\Ub^+ +
\Ub^-)/2$ and relative plate velocity $\vb = \Ub^+ - \Ub^-$, we can
express the solution as
\begin{equation}
  \ub = \overline{\Ub} + \left[ \Ab_s + \left( \Bb_s + \Ib \right)
    \frac{2z}{h} \right]   \frac{\vb}{2}  \label{eq:shear}
\end{equation}
where $\Ab_s$ and $\Bb_s$ are dimensionless $2\times 2$ matrices with
the following physical interpretations. The first term in
(\ref{eq:shear}) describes a slip-driven plug flow in the $\Ab_s \vb$
direction, and the second describes a slip-driven linear shear flow in
the $\Bb_s \vb$ direction. Substituting (\ref{eq:shear})
into (\ref{eq:BC}), we find
\begin{equation}
  \Ab_s = - \Db (\Ib + \Cb)^{-1} \ \ \mbox{ and } \ \ \Bb = (\Ib +
  \Cb)^{-1} - \Ib,   \label{eq:AB_s} 
\end{equation}
where
\begin{equation}
  h \Cb = \bb^+ + \bb^- \ \ \mbox{ and } \ \ 
  h \Db = \bb^+ - \bb^- .   \label{eq:CD}
\end{equation}
The slip-driven plug flow vanishes ($\Ab_s=\Ob$) only if the textures
are the same ($\Db = \Ob$), and slip-driven shear flow always occurs
($\Bb_s\neq \Ob$, if $\bb^+\neq 0$ or $\bb^-\neq 0$ and thus $\Cb \neq
\Ob$).

The solution (\ref{eq:shear})--(\ref{eq:AB_s}) exists for any
$\bb^\pm$, as long as $\Ib + \Cb$ is invertible; this is ensured for
passive surfaces, since $\bb^\pm$ and $\Cb$ are positive definite, and
possible for some active surfaces. In the typical case of symmetric
$\bb^\pm$, the solution can be expressed in terms of the texture
orientation angles $\theta^{\pm}$ and slip-length eigenvalues,
$b^{\pm}_\| = M^{\pm}_\| \eta$ and $b^{\pm}_\perp = M^{\pm}_\perp
\eta$ using (\ref{eq:Msym}). This can be easily accomplished in the
following general situations by diagonalizing $\Ab_s$ and $\Bb_s$.

\subsection{ Aligned but different textures }

We first consider `aligned' surfaces with the same orientation
$\theta^\pm = \theta$, but arbitrary slip-length eigenvalues:
\begin{equation}
\bb^\pm = \Sb_\theta \left( \begin{array}{cc}
      b^\pm_\| & 0 \\
      0 & b^\pm_\perp \end{array} \right) \Sb_{-\theta}   \label{eq:baligned}
\end{equation}
The coefficient tensors (\ref{eq:AB_s}) are then
diagonalized by the same rotation matrix
\begin{eqnarray}
\Ab_s &=& \Sb_\theta \left( \begin{array}{cc}
      A_s(b^+_\|,b^-_\|) & 0 \\
      0 & A_s(b^+_\perp,b^-_\perp)  \end{array} \right) \Sb_{-\theta} \\
\Bb_s &=& \Sb_\theta \left( \begin{array}{cc}
      B_s(b^+_\|,b^-_\|) & 0 \\
      0 & B_s(b^+_\perp,b^-_\perp)  \end{array} \right) \Sb_{-\theta}
\end{eqnarray}
and the eigenvalues
\begin{equation}
A_s(b^+,b^-) = -\frac{b^+ + b^-}{h + b^+ + b^-} \ \ \mbox{ and } \ \  
B_s(b^+,b^-) = -\frac{b^+ - b^-}{h + b^+ + b^-}
\end{equation}
result from scalar slip in the eigendirections.

\subsection{ Identical but misaligned textures }

Next we consider identical textures with arbitrary orientations,
$\theta^\pm = \bar{\theta} \pm \Delta \theta$:
\begin{equation}
\bb^\pm = \Sb_{\theta^\pm} \left( \begin{array}{cc}
      b_\| & 0 \\
      0 & b_\perp \end{array} \right) \Sb_{-\theta^\pm}   \label{eq:btilted}
\end{equation}
By expressing the sum and difference matrices (\ref{eq:CD}) as
\begin{eqnarray}
  h \Cb &=& 2 \Sb_{\bar{\theta}}  \left( \begin{array}{cc}
      b_\|\cos^2\Delta\theta + b_\perp\sin^2\Delta\theta & 0 \\
      0 &  b_\|\sin^2\Delta\theta + b_\perp\cos^2\Delta\theta 
      \end{array} \right)  \Sb_{-\bar{\theta}}   \label{eq:Ctilt} \\   
  h \Db &=& \Sb_{\bar{\theta}}  \left( \begin{array}{cc}
      0 & (b_\| - b_\perp) \sin 2\Delta\theta \\
      (b_\| - b_\perp) \sin 2\Delta\theta  & 0 \end{array} \right)  
      \Sb_{-\bar{\theta}}   \label{eq:Dtilt}
\end{eqnarray}
we find $\Ab_s = \Ob$ and
\begin{equation}
\Bb_s = \Sb_{\bar{\theta}}  \left( \begin{array}{cc}
      -\frac{b_\|\cos^2\Delta\theta + b_\perp\sin^2\Delta\theta}
           {h/2 + b_\|\cos^2\Delta\theta + b_\perp\sin^2\Delta\theta} & 0 \\
      0 & -\frac{b_\|\sin^2\Delta\theta + b_\perp\cos^2\Delta\theta}
           {h/2 + b_\|\sin^2\Delta\theta + b_\perp\cos^2\Delta\theta} 
      \end{array} \right)  \Sb_{-\bar{\theta}} .
\end{equation}
The slip-driven plug flow vanishes by symmetry , and the slip-driven
shear flow coefficient $\Bb_s$ is diagonalized by
$\Sb_{\bar{\theta}}$, where $\bar{\theta}$ is the angle that bisects
the orientation angles. As expected by symmetry, if the two textures
are the same (but misaligned), then shearing in this direction cannot
produce any transverse flow.

\section{ Example: Pressure-driven flow }
\label{sec:pois}

\subsection{ General solution }

Another simple solution to (\ref{eq:NS})--(\ref{eq:BC}) describes
steady, laminar flow in response to an applied pressure gradient, $\gb
= -\del p = g_x \xhat + g_y \yhat$, between stationary textured plates
in Fig.~\ref{fig:plates}. We express the solution in the form
\begin{equation}
  \ub = \frac{h^2}{4\eta} \left\{  \frac{1}{2} \left[ 1 -
        \left(\frac{2z}{h}\right)^2\right] \, \Ib + \Ab_p + \left(
          \frac{2z}{h} \right) \, \Bb_p   \right\} \, \gb \label{eq:pois}
\end{equation}
where $\Ab_p$ and $\Bb_p$ are dimensionless $2\times 2$ matrices. In
spite of surface anisotropy, the velocity is horizontal ($\zhat \cdot
\ub = 0$) and varies only in the vertical $z$ direction, due to
translational invariance. The solution (\ref{eq:pois}) is a linear
superposition of three terms: The first is the familiar parabolic
profile of Poiseuille flow in the $\gb$ direction between parallel
no-slip planes; the second is a slip-driven plug flow in the $\Ab_p
\gb$ direction; the third is a linear shear flow in the $\Bb_p \gb$
direction, which arises only if $\bb^+ \neq \bb^-$. Substituting
(\ref{eq:pois}) into (\ref{eq:BC}), we find
\begin{equation}
\Ab_p = \Cb - \Db (\Ib + \Cb )^{-1} \Db  \ \ \mbox{ and } \ \ 
\Bb_p = (\Ib + \Cb)^{-1} \Db   \label{eq:AB}
\end{equation}
in terms of the sum and difference tensors defined in (\ref{eq:CD}).

In the limit of no slip on the upper surface $\bb^+=\Ob$, our solution
reduces to that of \cite{stone2004}. In that case, $h \Cb = - h \Db =
\bb^-$, the coefficient tensors, $\Ab_p$ and $\Bb_p$, and the
permeability $\Kb$ are all coaxial with the slip-length tensor $\bb^-$
of the lower surface.  Here, we analyze more general situations where
the upper and lower surfaces have different slip tensors. 

For symmetric $\bb^\pm$, we can diagonalize $\Ab_p$ and $\Bb_p$ in the
same simple situations considered above for shear flow. In the case of
aligned but different slip tensors (\ref{eq:baligned}), the
coefficient tensors (\ref{eq:AB}) are diagonalized by the
same rotation matrix:
\begin{eqnarray}
\Ab_p &=& \Sb_\theta \left( \begin{array}{cc}
      A_s(b^+_\|,b^-_\|) & 0 \\
      0 & A_s(b^+_\perp,b^-_\perp)  \end{array} \right) \Sb_{-\theta} \\
\Bb_p &=& \Sb_\theta \left( \begin{array}{cc}
      B_s(b^+_\|,b^-_\|) & 0 \\
      0 & B_s(b^+_\perp,b^-_\perp)  \end{array} \right) \Sb_{-\theta}
\end{eqnarray}
and the eigenvalues
\begin{equation}
A_s(b^+,b^-) = \frac{b^+ + b^- + 4 h^{-1} b^+ b^- }{h + b^+ + b^- } \
\ \mbox{ and } \ \ 
B_s(b^+,b^-) = \frac{b^+ - b^-}{h + b^+ + b^- }
\end{equation}
result from scalar slip in the eigendirections. There are several
simple cases: (i) If the surfaces are isotropic, $\bb^\pm = b^\pm
\Ib$, then $\Ab_p = A_s \Ib$ and $\Bb_p = B_s \Ib$; (ii) if the
surfaces have the same slip tensors, $\bb^+ = \bb^- = \bb$, then
$\Ab_p = 2h^{-1} \bb$ and $\Bb_p = \Ob$; (iii) If the upper surface
has no slip, $\bb^+ = \Ob$ and $\bb^-=\bb$, then $A_s = -B_s = b / (h
+ b)$, or more compactly $\Ab_p = -\Bb_p = \bb (h \Ib + \bb)^{-1}$,
which reduces our solution to that of \cite{stone2004} for one
textured surface.

For identical but misaligned textures (\ref{eq:btilted}), we find
$\Ab_p = \Cb$ and $\Bb_p = \Ob$, using (\ref{eq:Ctilt}) and
(\ref{eq:Dtilt}). Now the slip-driven shear flow vanishes by
symmetry. The slip-driven plug flow is proportional to the average
slip-length tensor and diagonalized by $\Sb_{\bar{\theta}}$, where
$\bar{\theta}$ is the angle that bisects the surface orientation
angles. As expected, a pressure gradient in this direction cannot
produce any transverse flow, if the two textures are the same.

\subsection{ Permeability  }

In many situations, one is more interested in the depth-integrated
total flow rate in a given direction, rather than the velocity
profile. In linear response, the depth-averaged velocity
$\overline{\ub}$ is proportional to the applied pressure gradient,
\begin{equation}
\overline{\ub} = \frac{1}{h} \int_{-h/2}^{h/2} \ub dz = \kappab\,
\gb.  \label{eq:kdef}
\end{equation}
via the permeability tensor $\kappab$. For the anisotropic Poiseuille
flow (\ref{eq:pois}), this integral is easily performed to obtain
\begin{equation}
\kappab = \frac{h^2}{12 \eta} \Kb, \ \ \mbox{ where } \ \
 \Kb = \Ib + 3 \Ab_p   \label{eq:K}
\end{equation}
is the dimensionless permeability, scaled to its value without
slip. The permeability is generally enhanced by slip-driven plug flow
in the direction $\Ab_p \gb$. (The slip-driven shear flow does not
affect the permeability, although it contributes to mixing and
dispersion.)

The results above for $\Ab_p$ in various special cases can be extended
to $\Kb$, since the two tensors are coaxial:
\begin{equation}
  \Kb = \Ib + 3 \Ab_p = \Sb_{\theta_K} \left( \begin{array}{cc}
      K_\| & 0 \\
      0 & K_\perp  \end{array} \right) \Sb_{-\theta_K} \\
\end{equation}
where $\Sb_{\theta_K}$ diagonalizes $\Ab_p$ and $\Kb$. For aligned but
different textures (\ref{eq:baligned}), the permeability clearly has
the same orientation as the textures, $\theta_K=\theta$, and its
eigenvalues, $K_\| = K_s(b^+_\|,b^-_\|)$ and $K_\perp =
K_s(b^+_\perp,b^-_\perp)$ correspond to analogous cases of scalar
slip,
\begin{equation}
K_s(b^+,b^-) = \frac{h + 4(b^+ + b^-) + 12 h^{-1} b^+ b^-}{h + b^+ +
  b^-}
\end{equation}
For identical but misaligned textures (\ref{eq:btilted}), the
permeability is orientated with the mean angle $\theta_K=\bar{\theta}
= (\theta^+ + \theta^-)/2$ with eigenvalues given by
\begin{eqnarray}
  K_\| &=& 1 + \frac{6}{h} \left( b_\| \cos^2\Delta\theta
    + b_\perp \sin^2\Delta\theta \right) \\
  K_\perp &=& 1 + \frac{6}{h} \left( b_\| \sin^2\Delta\theta
    + b_\perp \cos^2\Delta\theta \right)
\end{eqnarray}
where $\Delta\theta = (\theta^+ - \theta^-)/2$. If
$\Delta\theta=\pi/4$, then the permeability is isotropic, $\Kb = K
\Ib$ with $K = K_\| = K_\perp = 1 + (3/h)(b_\| + b_\perp)$.

Microfluidic devices often contain thin channels of rectangular cross
section with parallel side walls at $y = \pm L$ with $L \gg h$. In
that case, the mean downstream permeability of the channel,
$\tilde{\kappa}_x=(h^2/12\eta) \tilde{K}_x$, defined by
$\overline{u}_x = \tilde{\kappa}_x g_x$, can be easily derived from
the permeability tensor $\kappab$ defined in 
Eq.~(\ref{eq:kdef}). Ignoring departures from Poiseuille flow within
$O(h)$ of the side walls, the constraint of vanishing transverse flow,
$\overline{u}_y=0$, is maintained by an induced transverse pressure
gradient, $g_y = -(\kappa_{yx}/\kappa_{yy}) g_x$, which drives an
additional anisotropic Poiseuille flow. Superimposing these flows, we
obtain
\begin{equation}
\tilde{K}_x = K_{xx} - \frac{K_{xy}K_{yx}}{K_{yy}}= 
\frac{\mbox{det}(\Kb) }{K_{yy}} = \frac{ K_\| K_\perp }{ K_{yy} }
\end{equation}
The channel permeability can also be interpretted in terms of an
effective downstream slip length $\tilde{b}_x$ defined by $\tilde{K}_x
= 1 + (6/h) \tilde{b}_x$, although this obscures the true tensorial
nature of the hydrodynamic slip.

\section{ Conclusion }
\label{sec:disc}

Our solutions for anisotropic flows between textured plates may be
useful in interpretting experiments and simulations. As in the case of
scalar Poiseille flow, bulk velocity profiles can be fitted to the
theory to systematically extract boundary effects of slippage and
assess the validity of the tensorial slip hypothesis. Our results also
allow the local slip tensors to be determined by global measurements,
such as the permeability of a textured channel or the force required
to shear textured plates, as a function of the surface
orientations. In such measurements, departures from our predictions
could be used to isolate nonlinear, inhomogeneous, or non-symmetric
slip response, e.g. due to nanobubble deformation at superhydrophobic
surface~\citep{sbragaglia.m:2007}, surface
curvature~\citep{vinogradova:95}, or variable channel
width~\citep{lauga2004}.

More generally, our calculations illustrate the power of the tensor
formalism to capture complicated effects of textured surfaces, while
preserving simple fluid domains. The general boundary condition
(\ref{eq:M}) may be useful for analytical or numerical calculations in
many other situations, such as lubrication flows between textured
gears, spreading or drainage of thin films, dispersion and mixing in
grooved channels~\citep{stroock2002a,stroock2002b}, sedimention of
textured particles~\citep{lecoq.n:2004}, and electrokinetics of
patterned surfaces~\citep{ajdari2002}. Transverse spatial couplings
could also be added to existing tensorial (but isotropic) slip
boundary conditions for fluids with internal degrees of freedom, such
as liquid crystals and polymer melts; anisotropic surface texture can
influence molecular orientations and thus the effective
slip~\citep{heidenreich2007}, which could have interesting
consequences for theory and applications.

\medskip

The authors gratefully acknowledge the hospitality of ESPCI and
support by the Paris-Sciences Chair (MZB) and Joliot Chair (OIV).
\bibliography{slip2} \bibliographystyle{jfm}

\end{document}